\documentclass[11pt]{article}
\usepackage{epsfig} 
\newcommand{\half}{\mbox{\small{$\frac{1}{2}$}}} 
 
\newcommand{\Nf}{N_{\!f}} 
\newcommand{\NA}{N_{\!A}} 
\newcommand{\MSbar}{\overline{\mbox{MS}}} 

\begin{document}
\title{One loop renormalization of the non-local gauge invariant operator 
$\stackrel{\mbox{\begin{small}min\end{small}}}
{\mbox{\begin{tiny}$\{U\}$\end{tiny}}} \int d^4x \, ( A^{a \, U}_\mu )^2$ 
in QCD} 
\author{J.A. Gracey, \\ Theoretical Physics Division, \\ 
Department of Mathematical Sciences, \\ University of Liverpool, \\ P.O. Box 
147, \\ Liverpool, \\ L69 3BX, \\ United Kingdom.} 
\date{} 
\maketitle 
\vspace{5cm} 
\noindent 
{\bf Abstract.} We compute the one loop anomalous dimension of the gauge
invariant dimension two operator $\stackrel{\mbox{\begin{small}min\end{small}}}
{\mbox{\begin{tiny}$\{U\}$\end{tiny}}} \int d^4x \, ( A^{a \, U}_\mu )^2$,
where $U$ is an element of the gauge group, by exploiting Zwanziger's expansion
of the operator in terms of gauge invariant non-local $n$ leg operators. The 
computation is performed in an arbitrary linear covariant gauge and the 
cancellation of the gauge parameter in the final anomalous dimension is 
demonstrated explicitly. The result is equivalent to the one loop anomalous 
dimension of the local dimension two operator $(A^a_\mu)^2$ in the Landau 
gauge. 

\vspace{-18cm}
\hspace{13.5cm}
{\bf LTH 745} 

\newpage 

The last decade has witnessed an intense interest in the area of dynamical
gluon mass generation related to the understanding of the low energy properties
of Quantum Chromodynamics (QCD). See, for instance, 
\cite{1,2,3,4,5,6,7,8,9,10,11}. One motivation for this has been the numerical
evidence for the apparent condensation of a dimension two operator to explain 
the deviation of an effective coupling constant from the expected perturbation 
theory prediction, \cite{12,13,14,15}. To reconcile the difference one can fit 
the data more accurately with a $1/Q^2$ correction requiring a dimension two 
operator to balance the dimensionality of the momentum $Q$. Ordinarily one 
would expect a dimension four operator correction as the leading power 
correction due to the gauge invariant operator $( G^a_{\mu\nu} )^2$ based on 
the field strength $G^a_{\mu\nu}$ where $a$ is the adjoint colour index. That a
dimension two operator appears to emerge in this analysis, \cite{12,13,14,15}, 
is not inconsistent with a variety of other observations made over a period of 
years. Indeed \cite{16,17,18} noted that the perturbative vacuum is unstable 
and the condensation of a dimension two operator would be energetically 
favourable. Earlier studies of potential gluon mass operators included a 
Coulomb gauge analysis of a dimension two operator, \cite{19}, as well as 
Cornwall's construction of a massive gauge invariant QCD Lagrangian which 
supports vortex solutions, \cite{20,21}. However, one main theoretical 
objection to such dimension two operators is that the obvious naive choice, 
$\half ( A^a_\mu )^2$, where $A^a_\mu$ is the gluon field, is clearly gauge 
variant and therefore not suitable for condensing in quantities involving gauge 
invariant objects. As the effective coupling constant of \cite{12,13,14,15} is 
gauge dependent, there is therefore no immediate reason why such a gauge 
variant operator condensate cannot be the explanation of the $O(1/Q^2)$ power 
correction. However, phenomenological analyses of gauge invariant quantities, 
\cite{1}, appear to require a gluon mass, albeit tachyonic, to fit experimental
data. Motivated by Curci and Ferrari's work from the 1970's, \cite{22}, there 
has been a recent re-examination of $\half ( A^a_\mu )^2$ and its BRST 
invariant extension since their model possesses a classical gluon mass term. 
The main drawback of the Curci-Ferrari model for being a possible Lagrangian of
a massive vector boson is that the BRST operator in not nilpotent and hence
unitarity is absent, \cite{23,24}. Instead the hope would be that the 
{\em quantum} condensation of the operator would circumvent the unitarity 
objection. Though this is still an open question. 

In this respect one calculation of note was the Landau gauge analysis of 
\cite{4} for the condensation of $\half ( A^a_\mu )^2$ in Yang-Mills theory 
using the local composite operator method. This was later extended to QCD for 
massless quarks in \cite{25}. The premise of investigation of \cite{4} rests in
the observation that one can in fact have a gauge invariant dimension two 
operator in QCD which can condense. Indeed the resultant phenomenology of the 
particular operator in question has been elaborated on in \cite{26}. This 
operator is given by $\stackrel{\mbox{\begin{small}min\end{small}}} 
{\mbox{\begin{tiny}$\{U\}$\end{tiny}}} \int d^4x \, ( A^{a \, U}_\mu )^2$,
where $U$ is an element of the gauge group which transforms the gauge field 
along a gauge orbit, and is by nature non-local. Its role in constructing a 
gauge fixing which is globally consistent and devoid of Gribov ambiguities, 
\cite{27}, has been discussed in, for instance, \cite{28,29,30,31}. That 
non-locality should play a role in aiming to describe infrared gluon dynamics 
should come as no surprise in that asymptotic freedom indicates that only at 
high energies are quarks and gluons effectively free whilst being hard to 
separate at low energies with lower energy interactions needing to be 
communicated over large distances. For \cite{4} the main initial technical 
hurdle to be overcome was the fact that the non-local operator is an infinite 
coupling constant series in an arbitrary gauge. Hence to do a full perturbative
analysis and construct a gauge invariant effective potential was initially 
impossible. However, by taking the point of view that such an effective 
potential exists then it seemed sensible to consider it in one gauge. 
Specifically, the Landau gauge was chosen whence the gauge invariant non-local 
operator truncates to the one local term $\half ( A^a_\mu )^2$, \cite{4}. 
Moreover, this operator is renormalizable leading to the successful analysis of
the operator's condensation in the two loop effective potential, \cite{4,25}. 
Interestingly, the operator in the Landau gauge does not possess an independent 
renormalization since its anomalous dimension is the sum of the gluon and 
Faddeev-Popov ghost anomalous dimensions, \cite{32,33,34}. A similar structure 
is present in the analogous operator in the Curci-Ferrari gauge, \cite{35,36}, 
and the maximal abelian gauge, \cite{37,38}. 

In light of these observations one would still wish to handle the gauge
invariant non-local operator itself within the context of quantum field theory
with the ultimate aim of constructing a gauge independent effective potential 
in order to study the condensation, similar to \cite{4,25}. As a first stage in 
such an exercise, the main purpose of this article is to renormalize  
$\stackrel{\mbox{\begin{small}min\end{small}}}
{\mbox{\begin{tiny}$\{U\}$\end{tiny}}} \int d^4x \, ( A^{a \, U}_\mu )^2$ at
one loop in an arbitrary covariant gauge and verify its independence of the 
gauge parameter as well as the equivalence of its anomalous dimension to the 
Landau gauge expression. From the nature of the operator this is not a trivial 
task as one has to renormalize a non-local operator in a gluon $2$-point 
function. Interestingly it transpires that a part of the calculation has 
already been performed but in a different context, \cite{39,40}, which we 
naturally adapt for the present work. Before we discuss this it is necessary to
indicate the relevant background to the problem and indicate the pitfalls of an
initial naive approach. First, we set notation for building the gauge invariant 
non-local operator by defining
\begin{equation}
{\cal O} ~ \equiv ~ \frac{1}{2} \stackrel{\mbox{\begin{small}min\end{small}}}
{\mbox{\begin{tiny}$\{U\}$\end{tiny}}} \int d^4x \, \left( A^{a \, U}_\mu 
\right)^2 ~. 
\end{equation}
Given a gauge field $A_\mu$~$=$~$A^a_\mu T^a$, where $T^a$ are the colour group 
generators, it is transported along a gauge orbit by the (global) gauge
transformation 
\begin{equation}
A^U_\mu ~=~ U A_\mu U^\dagger ~-~ \frac{i}{g} \left( \partial_\mu U \right) 
U^\dagger 
\end{equation} 
where $g$ is the coupling constant. By construction, the field $A^U_\mu$ is 
gauge invariant and hence it is trivial to see that the operator ${\cal O}$ is 
gauge invariant. However, the explicit form of $A^U_\mu$ can be determined by 
setting
\begin{equation}
U ~=~ e^{i \phi^a T^a}
\end{equation}
where $\phi^a$ can be deduced order by order in perturbation theory. For 
instance, \cite{30,39},
\begin{eqnarray}
A^{a \,U}_\mu &=& \left[ \eta_{\mu\nu} - 
\frac{\partial_\mu \partial_\nu}{\partial^2} \right] 
\left[ A^{a \, \nu} ~+~ g f^{abc} \left( \frac{1}{\partial^2} 
\partial^\sigma A^b_\sigma \right) A^{c \, \nu} \right. \nonumber \\
&& \left. ~~~~~~~~~~~~~~~~~~~-~ \frac{g}{2} f^{abc} 
\left( \frac{1}{\partial^2} \partial^\sigma A^b_\sigma \right) 
\left( \frac{1}{\partial^2} \partial^\nu \partial^\rho A^c_\rho \right) ~+~
O(g^2) \right]  
\end{eqnarray}  
where $f^{abc}$ are the colour group structure constants, from which it follows
that
\begin{eqnarray}
{\cal O} &=& \frac{1}{2} \int d^4 x \, \left[ A^a_\mu \left( \eta^{\mu\nu} - 
\frac{\partial^\mu\partial^\nu}{\partial^2} \right) A^a_\nu ~-~ 
2 g f^{abc} \left( \frac{1}{\partial^2} \partial^\nu \partial^\sigma A^a_\sigma 
\right) \left( \frac{1}{\partial^2} \partial^\rho A^b_\rho \right) A^c_\nu 
\right. \nonumber \\ 
&& \left. ~~~~~~~~~~~~-~ g f^{abc} A^a_\nu 
\left( \frac{1}{\partial^2} \partial^\sigma A^b_\sigma \right) 
\left( \frac{1}{\partial^2} \partial^\nu\partial^\rho A^c_\rho \right) ~+~ 
O(g^2) \right] ~.  
\label{aminexp} 
\end{eqnarray} 
In QED the operator truncates and represents the square of the transverse 
component of the gauge field. In the non-abelian case the operator involves an 
infinite number of terms and is assumed to converge. 

\begin{figure}[ht]
\hspace{6cm} 
\epsfig{file=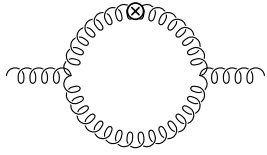,height=2cm}
\vspace{0.5cm}
\caption{Two leg operator insertion in gluon $2$-point function.} 
\end{figure} 
Naively one would expect to be able to renormalize this version of ${\cal O}$
order by order in perturbation theory since at any order there are only a 
finite number of terms. It transpires that this is not possible without
generating several difficult technical problems. First, at one loop one would
have to extract the divergent part of the topologies represented in figures
$1$ and $2$ where the encircled cross denotes the location of the operator
insertion. Ordinarily to renormalize the graph of figure $1$ one inserts the 
operator at zero momentum. However, as is well known, (see, for example, 
\cite{41}), one can obtain spurious results since the basis of operators into 
which the Feynman integral decomposes is not closed or complete for this
momentum configuration. Therefore, one has to have non-zero momentum operator 
insertion. Nullifying one of the remaining external momenta, though, results in
additional infrared divergences which need to be handled. To circumvent these 
difficulties one approach is to introduce a spurious infrared regularizing mass
which allows for the nullification of external momenta. Whilst this is in 
principle possible for local operators, for the particular operator we are 
concerned with it has an inherent non-locality which could lead to further 
difficulties at higher loops.

\begin{figure}[hb]
\hspace{3cm} 
\epsfig{file=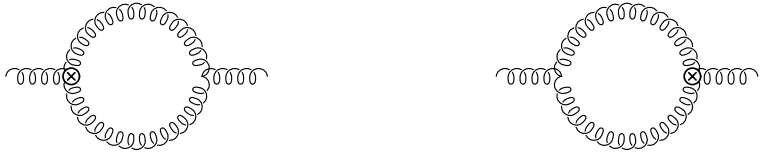,height=2cm}
\vspace{0.5cm}
\caption{Three leg operator insertion in gluon $2$-point function.} 
\end{figure} 
Rather than trying to handle these technical issues, it seems more appropriate
to regard ${\cal O}$ in a different way and use current results to extract its
anomalous dimension. In \cite{30} the gauge invariant operator ${\cal O}$ was 
rewritten as the sum of gauge invariant operators which can be treated 
individually in the renormalization procedure. Specifically, writing the 
summation of (\ref{aminexp}) in this non-perturbative way, \cite{30}, we have 
\begin{equation}
{\cal O} ~=~ \frac{1}{2} \int d^4 x \, \sum_{n=2}^\infty {\cal O}_n
\label{opexp} 
\end{equation}
where 
\begin{equation}
{\cal O}_2 ~=~ -~ \frac{1}{2} G^{a \, \mu\nu} \frac{1}{D^2} G^{a \, \mu\nu}
\label{op2}
\end{equation}
and, \cite{30},  
\begin{eqnarray}
{\cal O}_3 &=& g f^{abc} \left( \frac{1}{D^2} G^{a \, \mu\nu} \right) 
\left( \frac{1}{D^2} D^\sigma G^b_{\sigma\mu} \right) 
\left( \frac{1}{D^2} D^\rho G^c_{\rho\nu} \right) \nonumber \\  
&& -~ g f^{abc} \left( \frac{1}{D^2} G^{a \, \mu\nu} \right) 
\left( \frac{1}{D^2} D^\sigma G^b_{\sigma\rho} \right) 
\left( \frac{1}{D^2} D^\rho G^c_{\mu\nu} \right) 
\label{op3}
\end{eqnarray} 
with the covariant derivative, $D_\mu$, and field strength, $G^a_{\mu\nu}$,  
given by  
\begin{equation}
D_\mu A^a_\nu ~=~ \partial_\mu A^a_\nu ~-~ g f^{abc} A^b_\mu A^c_\nu 
~~~,~~~ 
G^a_{\mu\nu} ~=~ \partial_\mu A^a_\nu ~-~ \partial_\nu A^a_\mu ~-~
g f^{abc} A^b_\mu A^c_\nu ~. 
\end{equation}  
Indeed one could regard this as an expansion in operators with $n$ legs where
$n$ is determined from the lowest number gluon legs in each sub-operator
${\cal O}_n$.

The first term (\ref{op2}) of (\ref{opexp}) has already been studied in depth 
in \cite{39,40} where it was considered as a potential alternative mass 
operator for the gluon which was gauge invariant though non-local. Its 
anomalous dimension has been computed to two loops in the $\MSbar$ scheme in an
arbitrary linear covariant gauge and is given by, \cite{39,40},
\begin{eqnarray}
\gamma_{{\cal O}_2}(a) &=& -~ \frac{1}{3} \left[ 11 C_A - 4 T_F \Nf 
\right] a ~-~ \frac{\lambda^{abcd}\lambda^{abcd}}{128\NA} ~+~ f^{abe} f^{cde} 
\lambda^{adbc} \frac{a}{8\NA} \nonumber \\
&& +~ \left[ \frac{77}{12} C_A^2 - \frac{4}{3} C_A T_F \Nf - 4 C_F T_F \Nf
\right] a^2 ~+~ O(a^3)
\label{op2dim} 
\end{eqnarray}
in terms of $a$~$=$~$g^2/(16\pi^2)$ where 
\begin{equation}
\mbox{tr} ( T^a T^b ) ~=~ T_F \delta^{ab} ~~~,~~~ T^a T^a ~=~ C_F I ~~~,~~~
f^{acd} f^{bcd} ~=~ C_A \delta^{ab} 
\end{equation} 
and $\NA$ is the dimension of the adjoint representation. Clearly, the final 
expression is independent of the usual gauge parameter. To determine this 
result, the approach was to first localize ${\cal O}_2$ by introducing a set of 
additional localizing fields and associated (anti-commuting) ghost fields in 
such a way as to produce a renormalizable operator, \cite{39,40}. From the 
algebraic renormalizability analysis to ensure a multiplicatively 
renormalizable localization additional quartic interactions between all the 
localizing fields with couplings, $\lambda^{abcd}$, are required. These appear 
for the first time at two loops in the operator anomalous dimension, 
(\ref{op2dim}). Therefore, in the context of determining the anomalous 
dimension of ${\cal O}$ the contribution from the first term of (\ref{opexp}) 
is known. As we are only concerned with one loop, the piece involving the 
quartic couplings will not become relevant before any two loop renormalization. 
Hence, to complete the calculation of the anomalous dimension of ${\cal O}$, 
$\gamma_{{\cal O}}(a)$, all that is required is the piece deriving from 
(\ref{op3}) whose contributions will be deduced from the graphs of figure $2$. 
These give $\gamma_{{\cal O}_3}(a)$ whence $\gamma_{\cal O}(a)$ emerges at one 
loop from the sum of $\gamma_{{\cal O}_2}(a)$ and $\gamma_{{\cal O}_3}(a)$.

At one loop this is actually a simple calculation primarily as a result of the 
topology of the two graphs of figure $2$. Since the operator connects to an 
external leg, the problem of whether a zero or non-zero momentum actually flows 
through the operator insertion does not arise. The net flow through the
combination of external leg and connecting operator is non-zero which can be
distributed across both with neither being zero. In other words none of the 
earlier external momenta nullification complications arising in the two leg 
insertion of figure $1$ will arise for the graphs of figure $2$. Moreover, the 
non-locality resident in the operator does not lead to any additional 
difficulties and the Feynman rule is simple to derive. This is due to the fact 
that the operator $1/D^2$ can be replaced by $1/\partial^2$ at one loop and 
this only acts on {\em one} field for the leading leg term. Therefore, we have 
computed the two graphs of figure $2$ with ${\cal O}_3$ inserted in an 
arbitrary covariant gauge and found that the sum of the contributions to 
$\gamma_{{\cal O}}(a)$ from both graphs is 
\begin{equation}
\gamma_{{\cal O}_3}(a) ~=~ \frac{3}{4} C_A a ~+~ O(a^2)
\label{op3dim} 
\end{equation}
independent of the gauge parameter. For this particular calculation, we used
the {\sc Mincer} algorithm for the evaluation of massless $2$-point Feynman
diagrams, \cite{42,43}, written in the symbolic manipulation language 
{\sc Form}, \cite{44}, where the electronic representation of the graphs were 
generated via the {\sc Qgraf} package, \cite{45}, before being converted to 
{\sc Form} input notation. Moreover, we use dimensional regularization in
$d$~$=$~$4$~$-$~$2\epsilon$ dimensions and absorb the divergences into the 
renormalization constants using the $\MSbar$ scheme. The validity of using the
{\sc Mincer} algorithm for massless propagators follows because the Feynman
integrals are infrared safe and no infrared regularizing mass needs to be
introduced. The respective numerical contributions to (\ref{op3dim}) from the 
two operators of (\ref{op3}) are $33 C_A/4$ and $-$~$15 C_A/2$, which are each 
independent of the gauge parameter since each operator is itself gauge 
invariant. Thus adding all contributions, we find  
\begin{equation}
\gamma_{{\cal O}}(a) ~=~ -~ \frac{1}{12} \left[ 35 C_A - 16 T_F \Nf 
\right] a ~+~ O(a^2)
\end{equation}
where the independence of the gauge parameter is a trivial consequence. 
Moreover, the result is in agreement with the anomalous dimension of 
$\half ( A^a_\mu )^2$ computed in the Landau gauge, \cite{30,31,32}. 
Additionally the result appears to be consistent with the expectation that the 
anomalous dimension of a {\em local} gauge invariant operator is gauge 
independent, even though the operator itself is {\em non-local}.

We conclude by noting that it would appear that one can renormalize the
non-local operator ${\cal O}$ order by order in perturbation theory using
Zwanziger's non-perturbative decomposition into gauge invariant non-local
operators. However, a more stringent check of this would require a full 
{\em two} loop calculation which is a non-trivial exercise. Although the 
contribution to the overall anomalous dimension from the localized version of 
the two leg part of the operator is already available, \cite{39,40}, much of 
the groundwork has to be developed for the three and four leg operator 
insertions, such as the Feynman rules. Moreover, the four leg gauge invariant 
operator has yet to be constructed. In addition, in order to apply the 
{\sc Mincer} algorithm it is not inconceivable that both ${\cal O}_3$ and 
${\cal O}_4$ would need to be localized first by extending the original 
algebraic renormalization analysis of ${\cal O}_2$, \cite{39,40}. Nevertheless,
such a calculation would be interesting since it could validate the observation
that the full anomalous dimension of ${\cal O}$ is in fact given by the Landau 
gauge value which in  turn is determined from the sum of the gluon and 
Faddeev-Popov ghost anomalous dimensions in that specific gauge. 

\vspace{1cm} 
\noindent
{\bf Acknowledgement.} The author acknowledges useful discussions with D.J.
Broadhurst, R.E. Browne, D. Dudal, D. Kreimer, S.P. Sorella and H. Verschelde. 
Support from the Erwin Schr\"{o}dinger Institute, Vienna, Austria is also 
gratefully acknowledged as well as the organisers of the Mathematical and 
Physical Aspects of Perturbative Approaches to Quantum Field Theory programme 
during which part of this work was carried out.

\end{document}